\documentclass[article,showpacs,notitlepage]{revtex4-1}
\usepackage[pdftex]{color}
\usepackage{graphicx}
\usepackage{float}
\usepackage{subfigure}
\usepackage[applemac]{inputenc}
\usepackage[T1]{fontenc}
\usepackage{amssymb}
\usepackage{amsmath}
\usepackage{bm}
\usepackage{mathrsfs}
\usepackage{float}
\usepackage{slashed}
\usepackage[english]{babel}
\usepackage{comment}

\graphicspath{{./figurer/}}

\begin{document}

\title{Quantum kinetic theories in degenerate plasmas}
\author{Gert Brodin}
\author{Robin Ekman}
\author{Jens Zamanian}
\affiliation{Department of Physics, Ume{\aa } University, SE--901 87
Ume{\aa}, Sweden}

\begin{abstract}
In this review we give an overview of recent work on quantum kinetic
theories of plasmas. We focus, in particular, on the case where the
electrons are fully degenerate. For such systems, perturbation methods using
the distribution function can be problematic. Instead we present a model
that considers the dynamics of the Fermi surface. The advantage of this
model is that, even though the value of the distribution function can be
greatly perturbed outside the equilibrium Fermi surface, deformation of
Fermi surface is small up very large amplitudes. Next, we investigate the
short-scale dynamics for which the Wigner-Moyal equation replaces the Vlasov
equation. In particular, we study wave-particle interaction, and deduce that
new types of wave damping can occur due to the simultaneous absorption (or
emission) of multiple wave quanta. Finally, we consider exchange effects
within a quantum kinetic formalism to find a model more accurate than ones
using exchange potentials from density functional theory. We deduce the
exchange corrections to the dispersion relations for Langmuir and
ion-acoustic waves. Comparing with results based on exchange potentials
deduced from density functional theory we find that the latter models are
reasonably accurate for Langmuir waves, but rather inaccurate for ion
acoustic waves.
\end{abstract}

\maketitle

\section{Introduction}

There has been an increasing interest in plasmas of low-temperature and high densities, where quantum properties tend to be important.
A review of the recent evolution is given in e.g. Refs.~\cite{Haas-book,Manfredi,Shukla-Eliasson,Shukla-Eliasson-RMP}.
Promising applications include quantum wells~\cite{Manfredi-quantum-well}, spintronics~\cite{Spintronics} and plasmonics~\cite{Atwater-Plasmonics}.
Quantum plasma effects can also be of interest in experiments with solid density targets~\cite{glenzer-redmer}.
Important classifications of dense plasmas include whether they are strongly or weakly coupled, and whether they are degenerate or non-degenerate~\cite{Bonitz -1998}. 
While several works (see e.g. Refs.~\cite{Haas-book,Weak-coupling approx,Euruphys-2008,Spin-pond-1}) have applied quantum hydrodynamics, our focus here will be on the more accurate quantum kinetic theories \cite{zamanian2010kinet,Lundin2010,Manfredi-spin-kin,Andreev-PoP}.
Many familiar phenomena in plasma physics depend crucially on a kinetic description; these include 
wave-particle interaction, instabilities due to temperature anisotropy and finite Larmor radius effects. 
It can thus be expected that studying quantum kinetics will reveal new physics.

The cited theories generalize classical kinetic dynamics to include effects such as the spin, the Heisenberg uncertainty principle, particle-dispersive effects, degeneracy, and particle exchange effects.
New phenomena present in these theories include new wave modes~\cite{Brodin-et-al-PRL-2008}, a modification to the ponderomotive force due to spin-orbit interaction~\cite{Stefan-Brodin-2013}, and a new wave-damping process~\cite{ArXive-version} which we describe in Section~\ref{sec:waveDamping} of this paper. 
These theories can describe both long- and short-scale physics accurately in the non-relativistic regime.
In the semi-relativistic regime, a long-scale model exists~\cite{Asenjo}, and work on its fully relativistic generalization is ongoing~\cite{In prep}.

In the treatment here we will concentrate on the physics of fully degenerate electrons \cite{Alexandrov-book,Weak-c-dg1,Weak-c-dg2,Vladimirov-2012,Else-2010}, reviewing some recent findings.
However, it should be noted that many of the basic equations and methods (like e.g. the Wigner-Moyal equation~\cite{Shukla-Eliasson-RMP}) apply equally well to non-degenerate systems.
The organization of the paper is as follows. In Section~\ref{sec:degeneracy} we put forward a
quantum kinetic evolution equation that results from degeneracy alone \cite%
{Brodin-Srefan-PRE}. For a degenerate plasma perturbations of the
distribution function are never small compared to the background
distribution outside of the Fermi sphere. It is thus difficult to apply
standard perturbation theory. To circumvent this problem we instead study
the dynamics of the Fermi surface, which is only weakly perturbed. A
particular advantage with the approach outlined here is that it is
straightforward to generalize in order to cover spin dynamics, see Ref.~\cite%
{Brodin-Srefan-PRE}.

In Section~\ref{sec:waveDamping} we add the particle dispersive properties to the picture. Based
on the Schr\"{o}dinger equation we can then deduce the Wigner-Moyal equation.
This equation reduces to the classical Vlasov equation in the limit of long
scalelengths. The Wigner-Moyal equation is used to deduce a generalized
condition for wave-particle resonances. This resonance condition leads to a
new type of wave-particle damping \cite{ArXive-version}, corresponding to
the simultaneous absorption (or emission) of multiple wave quanta. This
mechanism may be the dominant one for plasmas of low temperature and high
densities.

In Section~\ref{sec:exchange} we investigate exchange dynamics. In particular we
make a comparison of expressions for exchange potentials based on density
functional theory (DFT) with results based on quantum kinetic theories \cite%
{Exchange.new}. A tentative conclusion is that the DFT potentials are
reasonably accurate for high-frequency dynamics when ions are immobile, but
less so when low-frequency phenomena is considered and the ion dynamics
comes into play. Finally, in Section~\ref{sec:conclusion} our results are discussed and
summarized.

\section{Fermi surface dynamics}
\label{sec:degeneracy}

As we will see in the next section, the Wigner-Moyal equation can be derived
without further approximations from the (single-particle) Schr\"{o}dinger
equation. Using the single-particle Schr\"{o}dinger equation naturally exchange
effects are neglected, which instead are covered in Section~\ref{sec:exchange}. For
macroscopic scale lengths longer than the characteristic de Broglie length,
the Wigner-Moyal reduces to the well-known Vlasov equation. Nevertheless, to
a certain extent, important quantum aspects can still be kept in the
modeling. By demanding that the background state of the electrons correspond
to an anti-symmetric many-body wave function, we replace the thermodynamic
background distribution with a Fermi-Dirac distribution. In the fully
degenerate limit with temperature $T=0$ this distribution reduces to%
\begin{equation}
f_{0}(\mathbf{v})=f_{\max }=\frac{m^{3}}{4\pi ^{3}\hbar ^{3}}  \label{Eq-1}
\end{equation}%
for $\left\vert \mathbf{v}\right\vert <v_{F}$ and $f_{0}=0$ otherwise. Here
the Fermi velocity is $v_{F}$ $=(\hbar /m_{e})(3\pi ^{2})^{1/3}n^{1/3}$,
where $n$ is the number density and $f_{\max }$ is the phase space density
maximally allowed by the Pauli exclusion principle. An important observation
here is that the classical dynamics respect the limitations imposed by the
Pauli exclusion principle. Propagating the particles along classical
trajectories, as implied by the Vlasov equation, preserves the phase space
density along particle orbits. Thus the magnitude of the distribution
function will never exceed the maximum value $f_{\max }$. \ For $T=0\,\ $we
instead have only two possible values of $f.$ Studying the fully degenerate
limit to some extent simplifies certain calculations, but applying weakly
nonlinear amplitude expansions become less straightforward. The reason is
that the division of $f=f_{0}+f_{1}$ where the perturbation fulfills $%
f_{1}\ll f_{0}$ becomes problematic when $f_{1}\neq 0$ in a phase space
region where $f_{0}=0$. If a particle is accelerated to a velocity slightly
above the Fermi velocity, this problem immediately occurs. A solution that
can be applied in such a scenario is to focus on the dynamics of the Fermi
surface. Since we know that $f=f_{\max }$ inside the Fermi surface, and $f=0$
outside, all quantities of interest can be computed given complete knowledge
of the evolution of the Fermi surface.

In order to derive an equation for the Fermi surface, we make the ansatz 
\begin{equation}
f=f_{\max }H(v-\tilde{v}(\mathbf{r},\phi _{v},\theta _{v},t)),  \label{Eq-2}
\end{equation}%
where $H$ is a step function with $H=1$ for negative argument and $H=0$
otherwise. Here $\phi _{v}$ and $\theta _{v}$ are the azimuthal angle and
polar angle in velocity space, respectively. Substituting this ansatz into
the Vlasov equation gives us%
\begin{equation}
\left( \frac{\partial }{\partial t}+\mathbf{v}\cdot \nabla +\frac{q}{m}%
\left( \mathbf{E}+\mathbf{v}\times \mathbf{B}\right) \cdot \nabla _{v\perp
}\right) \tilde{v}=\frac{qE_{r}}{m},  \label{Eq-3}
\end{equation}%
where $\mathbf{v=}\tilde{v}\mathbf{\hat{r}}_{v}$ and $\mathbf{\hat{r}}_{v}$
is a unit vector in the direction of the velocity, i.e. $\mathbf{\hat{r}}%
_{v}=(\sin \theta _{v}\cos \phi _{v},\sin \theta _{v}\sin \phi _{v},\cos
\theta _{v})$. $\nabla _{v\perp }$ is a velocity gradient perpendicular to $%
\mathbf{\hat{r}}_{v},$and $E_{r}=\mathbf{\hat{r}}_{v}\cdot \mathbf{E}$. To
have a closed system we need the source terms in Maxwell's equation in terms
of $\tilde{v}$. Using that $f=f_{\max }$ inside the Fermi surface, we
immediately find the charge density $\rho _{c}$ as 
\begin{equation}
\rho _{c}=qf_{\max }\int \frac{\tilde{v}^{3}}{3}d\Omega ,  \label{Eq-4}
\end{equation}%
and the current density $\mathbf{j}$ as

\bigskip 
\begin{equation}
\mathbf{j=}qf_{\max }\int \frac{\tilde{v}^{4}\mathbf{\hat{r}}_{v}}{4}d\Omega
,  \label{Eq-5}
\end{equation}%
where $d\Omega =\sin \theta _{v}d\phi _{v}d\theta _{v}$.

The theory outlined
here cannot be generalized to cover short scale physics, as the Wigner
function that generalizes the classical distribution function is not
conserved along particle orbits. However, to a certain degree it is possible
to cover dynamics involving the spin degrees of freedom using a generalized
version of Eq.~\eqref{Eq-3}. This is explored in some detail in Ref. \cite%
{Brodin-Srefan-PRE}. Furthermore, in that paper a solution of Eq.~\eqref{Eq-3} is computed for the case of nonlinear Landau damping, that
demonstrates the advantage of working with the dynamics of the Fermi
surface, rather than solving the full Vlasov equation. Specifically it is
shown that the Fermi surface is only weakly perturbed in the regime of
strongly nonlinear bounce oscillations (see Fig. 2 of Ref. \cite%
{Brodin-Srefan-PRE})

\section{Short-scale dynamics}
\label{sec:waveDamping}

For very short macroscopic scale lengths, wave function dispersion enters
the picture. In this case the classical Vlasov equation is replaced by the
Wigner-Moyal equation \cite%
{Haas-book,Manfredi,Shukla-Eliasson,Shukla-Eliasson-RMP}. A simple
derivation of the Wigner-Moyal equation for electrostatic fields can be made
by defining the Wigner function as 
\begin{equation}
f(\mathbf{r},\mathbf{v},t)=\frac{m^{3}}{(2\pi \hbar )^{3}}\int \psi ^{\ast }(%
\mathbf{r+r}^{\prime }/2,t)\psi (\mathbf{r-r}^{\prime }/2,t)\exp (im\mathbf{%
v\cdot r}^{\prime }/\hbar )d^{3}r^{\prime } ,  \label{Eq-6}
\end{equation}%
and computing the time evolution of $f$ based on the single-particle 
Schr\"{o}dinger equation 
\begin{equation}
i\hbar \frac{\partial \psi }{\partial t}+\frac{\hbar ^{2}}{2m}\nabla
^{2}\psi +q\Phi \psi =0 ,  \label{Eq-7}
\end{equation}%
where $\Phi $ is the electrostatic potential. It is then straightforward to
deduce the Wigner-Moyal equation 
\begin{equation}
\frac{\partial f}{\partial t}+\mathbf{v}\cdot \nabla f-\frac{qm^{3}}{\hbar }%
\int \frac{d^{3}r^{\prime }d^{3}v^{\prime }}{(2\pi \hbar )^{3}}\exp [im(%
\mathbf{v-v}^{\prime })\mathbf{\cdot r}^{\prime }/\hbar ][\Phi (\mathbf{r+r}%
^{\prime }/2,t)-\Phi (\mathbf{r-r}^{\prime }/2,t)]f(\mathbf{r},\mathbf{v}%
^{\prime },t)=0 .  \label{Eq-8}
\end{equation}
The connection of the Wigner-Weyl formalism to the algebra of
pseudo-differential operators is explored in Ref. \cite{Adj-1}. Furthermore,
the phase space analysis of wave equations presented in Ref. \cite{Adj-2} is
also highly relevant for the Wigner-Weyl formalism. As is well-known, unlike
the classical Vlasov equation, $f$ cannot be considered as a probability
density. In fact, $f$ can be negative in small regions of phase space.
Nevertheless, the charge and current density can be calculated in the same
manner as for a classical distribution function, i.e. 
\begin{equation}
\rho _{c}=q\int fd^{3}v  \label{Eq-9}
\end{equation}%
and 
\begin{equation}
\mathbf{j}=q\int \mathbf{v}fd^{3}v  \label{Eq-10c}
\end{equation}%
For\bigskip\ electrostatic fields only the charge density is needed and
Poisson's equation 
\begin{equation}
\nabla ^{2}\Phi =\frac{q}{\epsilon_0}\int fd^{3}v- \frac{q_{i} n_{0}}{%
\epsilon_0} ,  \label{Eq-11d}
\end{equation}%
together with \eqref{Eq-8} form a closed system. Here we have assumed a
constant neutralizing ion charge density $q_{i}n_{0}$, but if needed, it is
easy to relax this condition by including a dynamic model for the ions.

Before we discuss solutions to Eqs.~\eqref{Eq-8}~ and~\eqref{Eq-11d} let us
rewrite Eq.~\eqref{Eq-8} in an alternative way. Taylor expanding the
arguments of the potentials, doing multiple partial integrations, and using
a standard delta function relation, we obtain%
\begin{equation}
\frac{\partial f}{\partial t}+\mathbf{v}\cdot \nabla f-\frac{2q}{\hbar }\Phi
\sin \left( \frac{\hbar }{2m}\overleftarrow{\nabla }_{x}\cdot 
\overrightarrow{\nabla }_{v}\right) f(\mathbf{r},\mathbf{v},t)=0 .
\label{Eq-11.5}
\end{equation}%
Here the arrows indicate the direction the operators are acting, i.e. the
spatial gradient act on $\Phi $ and the velocity gradient act on $f$. The
sinus-operator is defined by its Taylor-expansion. By keeping just the first
order term in the expansion, we recover the classical Vlasov equation. In
this process the validity condition for dropping higher order terms is found
to be that the macroscopic scale lengths for variations in the potential are
much longer than the characteristic de Broglie length of the particles.

To illustrate a basic effect of the Wigner-Poisson system, let us study
small amplitude waves. Dividing the Wigner function as $f=F_{0}+f_{1}(%
\mathbf{v})\exp [i(kz-\omega t)]$ and linearizing, the solution for the
Wigner function is%
\begin{equation}
f_{1}=\frac{q\Phi \left[ F_{0}(\mathbf{v}+\mathbf{v}_{q})-F_{0}(\mathbf{v}-%
\mathbf{v}_{q})\right] }{\hbar (\omega -kv_{z})}.  \label{Eq-12}
\end{equation}%
where $\mathbf{v}_{q}=(\hbar k/2m)\mathbf{\hat{z}}$. Inserting this
expression into Poisson's equation, the linear dispersion relation becomes%
\begin{equation}
1=-\frac{q^{2}k^{2}}{\hbar \varepsilon _{0}}\int \frac{F_{0}(\mathbf{v}+%
\mathbf{v}_{q})-F_{0}(\mathbf{v}-\mathbf{v}_{q})}{(\omega -kv_{z})}d^{3}v.
\label{Eq-13}
\end{equation}%
Changing integration variables the dispersion relation can be written as 
\begin{equation}
1=-\frac{q^{2}k^{2}}{\hbar \varepsilon _{0}}\int \left( \frac{F_{0}(\mathbf{v%
})}{\omega -kv_{z}+\hbar k^{2}/2m}-\frac{F_{0}(\mathbf{v})}{\omega
-kv_{z}-\hbar k^{2}/2m}\right) d^{3}v.  \label{Eq-14}
\end{equation}%
This result has been studied in detail in e.g. Refs.~\cite%
{Shukla-Eliasson2009,Rightley2016}. The main effect we are interested in
here is the modification of the resonant velocity in the Landau poles. As is
apparent from Eq.~\eqref{Eq-14}, the resonant velocities are modified from
the classical case according to 
\begin{equation}
v_{\mathrm{res}}=\frac{\omega }{k}\rightarrow v_{\mathrm{res}}=\frac{\omega 
}{k}\pm \frac{\hbar k}{2m}.  \label{Eq-15}
\end{equation}%
Let us study the physical meaning of this modification. When a particle
absorbs or emits a wave quantum it can increase or decrease the momentum
according to 
\begin{equation}
\hbar k_{1}\pm \hbar k=\hbar k_{2},  \label{Eq-16}
\end{equation}%
and at the same time the energy changes according to%
\begin{equation}
\hbar \omega _{1}\pm \hbar \omega =\hbar \omega _{2}.  \label{Eq-17}
\end{equation}%
Next we identify $\hbar k_{1}/m$ (or equally well $\hbar k_{2}/m$) with the
resonant velocity $v_{\mathrm{res}}$ and note that for small amplitude waves
the particle frequencies and wavenumbers $(\omega _{1,2},k_{1,2})$ obey the
free particle dispersion relation $\omega _{1,2}=\hbar k_{1,2}^{2}/2m$.
Using these relations we see that the energy momentum relations Eqs.~\eqref{Eq-16}%
) and \eqref{Eq-17} imply the modification of the resonant velocity seen in
Eq.~(\eqref{Eq-15}). An interesting possibility, which was studied in Ref.~\cite{ArXive-version}, is the simultaneous absorption (or emission) of multiple wave
quanta, rather than a single wave quantum at a time. In that case Eqs.~\eqref{Eq-16} and \eqref{Eq-17} are replaced by 
\begin{equation}
\hbar k_{1}\pm n\hbar k=\hbar k_{2},  \label{Eq-18}
\end{equation}%
and 
\begin{equation}
\hbar \omega _{1}\pm n\hbar \omega =\hbar \omega _{2},  \label{Eq-19}
\end{equation}%
where $n=1,2,3, \ldots$ is an integer. Accordingly, the resonant velocities now
becomes%
\begin{equation}
v_{\mathrm{res}}=\frac{\omega }{k}\pm n\frac{\hbar k}{2m}.  \label{Eq-20}
\end{equation}%
When we pick the minus sign in Eq.~\eqref{Eq-20}, the resonant velocity for
absorbing multiple wave quanta can be considerably smaller, provided the
wavelengths are short. As a consequence, in the case of Langmuir waves, the
damping rate due to absorption of multiple wave quanta can be larger than
the standard linear damping rate. Basically this is due to a larger number
of resonant particles in the former case. These issues have been explored in
some detail in Ref. \cite{ArXive-version}, where the damping rates for
two-plasmon damping and three-plasmon damping have been computed.

\section{Exchange dynamics}
\label{sec:exchange}

The Wigner-Moyal equation studied in Section~\ref{sec:waveDamping} was derived from the
single-particle rather than the many-body version of the Schr\"{o}dinger
equation. As a result, it does not include exchange effects. Before we turn
to the quantum kinetic theories, let us consider an expression for exchange
effects that has been used rather extensively in a fluid formalism, see e.g.
Refs \cite{Weak-coupling approx,DFT-HF-LF,DFT-LF,DFT-rel,DFT-semicond}. Here
exchange potentials derived from density functional theory (DFT) \cite%
{Halperin} have been incorporated in a fluid formalism \cite{Weak-coupling
approx}. For one-dimensional spatial variations along $z$ the momentum
equation reads 
\begin{equation}
\frac{\partial u}{\partial t}+u\frac{\partial u}{\partial z}=\frac{q}{m}E+%
\frac{1}{m}\frac{\partial V_{x}}{\partial z}+\frac{\hbar ^{2}}{2m^{2}}\frac{%
\partial }{\partial z}\left( \frac{\partial ^{2}(\sqrt{n})/\partial z^{2}}{%
\sqrt{n}}\right) -\frac{1}{mn}\frac{\partial P}{\partial z}.  \label{Eq-21}
\end{equation}%
Here $u$ is the fluid velocity, $P$ is the fluid pressure, $n$ is the number
density and the third term of the right hand side is the Bohm de Broglie
potential that accounts for particle dispersive effects. The DFT exchange
potential is given by 
\begin{equation}
V_{x}=\frac{0.985\kappa }{4\pi }\frac{h^{2}\omega _{p}^{2}}{mv_{F}^{2}}%
\left( \frac{n}{n_{0}}\right) ^{1/3} ,  \label{Eq-22}
\end{equation}%
where $\kappa =(3\pi ^{2})^{2/3}$ and $n_{0}$ is the unperturbed number
density. Eqs.~\eqref{Eq-21} and \eqref{Eq-22} are complemented by the
continuity equation (same as in the classical case) and Poisson's equation.
A few things should be noted. Firstly, the exchange effects are often
presented along with a contribution from particle correlations (collisions).
The correlation contribution has been dropped here, as we would like to make
a comparison with the quantum kinetic exchange effects only. Secondly, the
derivation of the exchange potential has been made assuming a fully
degenerate plasma, i.e. the pressure $P$ in Eq.~\eqref{Eq-21} is the Fermi
pressure. Thirdly, we will be concerned with the long-scale limit (i.e.
characteristic wave-numbers $k$ that fulfill $\hbar k\ll mv_{F}$), in which
case the Bohm-de Broglie term can be neglected. Finally we note that the
sign of the exchange term is such as to counteract the pressure. It is worth
noting that there have been some confusion over the sign of the term in the
literature, but the original papers \cite{Weak-coupling approx,Halperin} has
the same (correct) sign as in Eq.~\eqref{Eq-21}.

While the DFT formalism can be derived from first principles, the use of
trial functions and approximations such as the adiabatic local density
approximation (ALDA) makes it important to verify the DFT potentials by
independent means. This can be made using a quantum kinetic formalism. By
writing down the first equation in the BBGKY-hierarchy and writing the
two-particle density matrix as a anti-symmetric product of one-particle
density matrices, a correction due to exchange effects can be obtained.
Assuming a plasma without spin polarization and summing over all spin
states, the following expression 
\begin{align}
& \partial _{t}f(\mathbf{x},\mathbf{p},t)+\frac{\mathbf{p}}{m}\cdot \nabla
_{x}f(\mathbf{x},\mathbf{p},t)+e\mathbf{E}(\mathbf{x},t)\cdot \nabla _{p}f(%
\mathbf{x},\mathbf{p},t)  \notag \\
& =\,\frac{1}{2}\partial _{p}^{i}\int d^{3}\!r\,d^{3}\!q\,\,e^{-i\mathbf{r}%
\cdot \mathbf{q}/\hbar }[\partial _{r}^{i}V(\mathbf{r})]f\left( \mathbf{x}-%
\frac{\mathbf{r}}{2},\mathbf{p}+\frac{\mathbf{q}}{2},t\right) f\left( 
\mathbf{x}-\frac{\mathbf{r}}{2},\mathbf{p}-\frac{\mathbf{q}}{2},t\right) 
\notag \\
& -\frac{i\hbar }{8}\partial _{p}^{i}\partial _{p}^{j}\cdot \int
d^{3}\!r\,d^{3}\!q\,\,e^{-i\mathbf{r}\cdot \mathbf{q}/\hbar }[\partial
_{r}^{i}V(\mathbf{r})]\left[ f\left( \mathbf{x}-\frac{\mathbf{r}}{2},\mathbf{%
p}-\frac{\mathbf{q}}{2},t\right) \left( \overleftarrow{\partial }_{x}^{j}-%
\overrightarrow{\partial }_{x}^{j}\right) f\left( \mathbf{x}-\frac{\mathbf{r}%
}{2},\mathbf{p}+\frac{\mathbf{q}}{2},t\right) \right] ,  \label{Eq-23}
\end{align}%
was derived in the long scale limit in Ref.\ \cite{Zamanian-exchange}. It
should be noted that Eq.~\eqref{Eq-23} is limited to electrostatic fields
(for a treatment allowing for electromagnetic fields, see Ref. \cite%
{Zamanian-II-exchange}). The long scale limit means that the macroscopic
scale lengths are longer than the characteristic de Broglie length, such
that the left hand side of Eq.~\eqref{Eq-23} corresponds to the Vlasov
limit. In the right hand side of Eq.~\eqref{Eq-23} we use $\mathbf{x}$ and $%
\mathbf{r}$ for position vectors and $\mathbf{p}$ and $\mathbf{q}$ for
momentum vectors. Furthermore, $\partial _{x}^{i}\equiv \partial /\partial
x_{i}$ and analogously for $\partial _{p}^{i}$ and $\partial _{r}^{i}$. An
arrow above an operator indicates in which direction it acts. We have also
used the summation convention so that a sum over indices occurring twice in
a term is understood. Finally $V(\mathbf{r})=$ $e^{2}/4\pi \varepsilon
_{0}\left\vert \mathbf{r}\right\vert $ is the Coulomb potential.

We are now interested in comparing the DFT predictions based on Eq.~\eqref{Eq-21}) with the quantum kinetic predictions based on Eq.~\eqref{Eq-23}. For this
purpose we consider the simple examples of linear Langmuir waves and linear
ion acoustic waves in homogeneous plasmas. We do the calculations
perturbatively, i.e. treating the exchange contribution as a small
correction. The background distribution of electrons is assumed to be fully
degenerate (see Eq.~\eqref{Eq-1}). For the case of Langmuir waves, the ions
are assumed to be immobile, and for the case of ion-acoustic waves, the ions
are treated classically and have a temperature $T=0$. The full details of
the calculations are presented in \cite{Exchange.new}. Here we just proceed
to the results. Starting with Langmuir waves for wavenumbers $k\ll \omega
/v_{F}$ (such that Landau damping do not occur) the result derived from Eq.~\eqref{Eq-23} is 
\begin{equation}
\omega ^{2}=\omega _{p}^{2}+\frac{3}{5}v_{\text{F}}^{2}k^{2}-\frac{3\hbar
^{2}\omega _{e}^{2}k^{2}}{20m_{e}^{2}v_{\text{F}}^{2}}.  \label{Eq-24}
\end{equation}%
where the last term is the exchange correction. This is in exact agreement
with previous calculations using several different methods, see Refs.~\cite%
{Roos-1961,Kanazawa-1960,Nozieres-1958}. However, Eq.~\eqref{Eq-23} as well
as the methods used in \cite{Roos-1961,Kanazawa-1960,Nozieres-1958} are
rather cumbersome to apply for more complicated problems (e.g. nonlinear
and/or inhomogeneous systems). Thus it is interesting to note that the
considerably simpler fluid formalism, based on Eq.~\eqref{Eq-21} is able to
give a comparatively good agreement. Replacing the numerical coefficient in
Eq.~\eqref{Eq-22} according to $0.985\rightarrow 1.23$ would give agreement
with Eq.~\eqref{Eq-24}. Now we turn to the case of ion-acoustic waves. The
dispersion relation derived from Eq.~\eqref{Eq-23} is then given by 
\begin{equation}
\omega ^{2}=\frac{m_{e}k^{2}v_{\text{F}}^{2}}{3m_{i}}\left[ 1-\frac{\hbar
^{2}\omega _{e}^{2}}{3m_{e}^{2}v_{\text{F}}^{4}}(14.9+7.11i)\right] ,
\label{Eq-25}
\end{equation}%
where the exchange correction coefficients $14.9$ and $7.11$ comes from
numerical solutions of certain integrals (see \cite{Exchange.new}).
Naturally the fluid formalism cannot produce the imaginary correction
corresponding to wave-particle interaction. However, comparing with the
dispersion relation based on Eq.~\eqref{Eq-21} it turns out that also the real
part of the dispersion relation is somewhat inaccurate. To get agreement, we
would need to modify the numerical coefficient in Eq.~\eqref{Eq-22}
according to $0.985\rightarrow 6.52$. A tentative conclusion is that the DFT
potential given in Eq.~\eqref{Eq-22} is reasonably accurate for high-frequency
phenomena involving only electrons, whereas it is inaccurate for
low-frequency phenomena involving also the ion dynamics. Such a conclusion
might be surprising, since it is usually more easy to obtain accurate DFT
approximations for time-independent or slowly varying phenomena. A possible
explanation is that wave-particle interaction is important for the
low-frequency regime, but not for the high-frequency regime. Hence a quantum
kinetic treatment is more crucial in the former case. However, more research
is needed before a definitive conclusion can be reached.

\section{Summary and discussion}
\label{sec:conclusion}

In order to have a clear focus, we have in this brief review concentrated on
fully degenerate systems, and in some cases also limited ourselves to
electrostatic fields. Moreover, we have neglected all effects due to spin
polarized systems (e.g. a magnetic dipole force due to the spin, spin
magnetization currents, spin precession, etc.) and also neglected
relativistic effects. However, most of the theories presented here can be
generalized to cover more general cases. Examples of works that discuss
quantum kinetic theories of spin polarized plasmas are Refs.~\cite%
{zamanian2010kinet,Lundin2010,Manfredi-spin-kin,Andreev-PoP}. Furthermore,
weakly \cite{Asenjo} and strongly \cite%
{Liboff-book,Haas-relativistic,Mendonca-rel} quantum relativistic effects
effects have also been studied. Of particular interest is a strongly quantum
relativistic treatment that also cover the spin dynamics \cite{In prep}

The conclusions from the present paper are as follows. Firstly we note that
the analysis of fully degenerate systems can be much simplified by studying
the dynamics of the Fermi surface, as described by Eq.~\eqref{Eq-3}--\eqref{Eq-5}. 
Secondly, we note that replacing the classical Vlasov equation with
the Wigner-Moyal equation modifies wave-particle interaction considerably.
As is well-known, the usual Landau poles have a \ velocity shift $\pm \hbar
k/2m$. Moreover, we deduce that new poles are produced when a nonlinear
analysis is made. This is associated with multi-plasmon damping, as
discussed in more detail in Ref. \cite{ArXive-version}. Thirdly, we have
used quantum kinetic theories of exchange effects, in order to evaluate the
accuracy of exchange potentials based on the DFT formalism. We have found
that the DFT potentials give reasonably correct results for the dispersion
relation of Langmuir waves. By contrast, the accuracy when it comes to the
ion acoustic dispersion relation is very low \cite{Exchange.new}. It is
suggestive to think that this imply a more general conclusion, i.e. that the
DFT potentials are accurate when pure electron motion is considered, but not
when the ion dynamics is involved. However, more research is needed to see
if such a conclusion is justified.

\end{document}